\documentclass[journal]{IEEEtran}
\usepackage{graphicx}
\usepackage{xcolor}
\usepackage{colortbl}
\usepackage{amssymb}
\usepackage{amsmath}
\usepackage{multirow}
\usepackage{booktabs}

\allowdisplaybreaks

\usepackage{cite}
\ifCLASSINFOpdf
\else
\fi
\usepackage{amsmath}
\begin{document}

% ---------- Preprint Notice ----------
\onecolumn
\noindent
\textbf{Preprint Notice:} This is a preprint of an article submitted to IEEE for possible publication. \\
\copyright~[2025] [Ahmad Mustafa]. This manuscript has not been peer reviewed or accepted yet. \\
Please do not redistribute without permission of the authors.

\vspace{1cm}
\twocolumn
% -------------------------------------

\title{ReCoGNet: Recurrent Context-Guided Network for 3D MRI Prostate Segmentation}

% \author{Ahmad Mustafa, Reza Rastegar, and Ghassan AlRegib% <-this % stops a space
% }

\author{
    \IEEEauthorblockN{Ahmad Mustafa,~\IEEEmembership{Member,~IEEE}, 
                      Reza Rastegar, and Ghassan AlRegib,~\IEEEmembership{Fellow,~IEEE}}
\thanks{Ahmad Mustafa (ahmadmustafa.am@gmail.com) and Ghassan AlRegib (alregib@gatech.edu) are affiliated to the School of Electrical and Computer Engineering, Georgia Institute of Technology, Atlanta, USA.}
}


\markboth{Journal of \LaTeX\ Class Files,~Vol.~14, No.~8, August~2015}%
{Shell \MakeLowercase{\textit{et al.}}: Bare Demo of IEEEtran.cls for IEEE Journals}



\maketitle

\begin{abstract}
Prostate gland segmentation from T2-weighted MRI is a critical yet challenging task in clinical prostate cancer assessment. While deep learning-based methods have significantly advanced automated segmentation, most conventional approaches—particularly 2D convolutional neural networks (CNNs)—fail to leverage inter-slice anatomical continuity, limiting their accuracy and robustness. Fully 3D models offer improved spatial coherence but require large amounts of annotated data, which is often impractical in clinical settings. To address these limitations, we propose a hybrid architecture that models MRI sequences as spatiotemporal data. Our method uses a deep, pretrained DeepLabV3 backbone to extract high-level semantic features from each MRI slice and a recurrent convolutional head, built with ConvLSTM layers, to integrate information across slices while preserving spatial structure. This combination enables context-aware segmentation with improved consistency, particularly in data-limited and noisy imaging conditions. We evaluate our method on the PROMISE12 benchmark under both clean and contrast-degraded test settings. Compared to state-of-the-art 2D and 3D segmentation models, our approach demonstrates superior performance in terms of precision, recall, Intersection over Union (IoU), and Dice Similarity Coefficient (DSC), highlighting its potential for robust clinical deployment.

 \end{abstract}


% \fi
\IEEEpeerreviewmaketitle

\section{Introduction}
\IEEEPARstart{P}{rostate} cancer is the second most common cancer among men globally and remains a major cause of cancer-related morbidity and mortality \cite{Fassia2024}. Accurate segmentation of the prostate gland from Magnetic Resonance Imaging (MRI), particularly T2-weighted sequences, plays a pivotal role in diagnosis, disease stratification, biopsy planning, and therapy monitoring \cite{Eddardaa2021}. However, manual annotation of prostate boundaries in MRI is time-consuming, prone to inter-observer variability, and difficult to scale across large datasets or clinical institutions. Automated segmentation using deep learning has emerged as a promising alternative, offering consistent and efficient delineations that support decision-making in clinical workflows \cite{Iosif2021}.

Convolutional neural networks (CNNs) have been widely adopted for this task due to their strong representational capabilities and success in other biomedical imaging domains \cite{Kuruparan2022, jensen2019prostate, khan2019zonal ,baldeon2020adaresu}. Most existing approaches, however, rely on 2D CNNs that process each MRI slice independently. While computationally efficient and less demanding in terms of labeled training data, these slice-wise models fail to capture inter-slice dependencies and anatomical continuity along the superior-inferior axis of the prostate. This can lead to segmentation inconsistencies, especially at the apex and base of the gland where tissue boundaries are more ambiguous.

Volumetric segmentation models based on 3D CNNs address this limitation by processing entire image stacks, thus incorporating contextual information across slices \cite{mehrtash2018automatic, to2018deep}. Yet, this gain in spatial coherence comes at the cost of higher computational complexity in 3D convolutional kernels and larger training data size, with substantial GPU resources required to train them effectively. These requirements are difficult to satisfy in real-world medical imaging settings, where data is often limited and heterogeneous in terms of acquisition protocols, scanners, and patient populations.

Recurrent neural networks (RNNs) and Long Short-Term Memory (LSTM) networks have long been used to model sequential dependencies in applications involving time series data such as speech recognition \cite{graves2013hybrid} and natural language processing \cite{wang2018lstm}. However, traditional recurrent architectures and their variants tend to not be as effective for spatiotemporal image applications requiring the prediction of dense outputs. This is because they flatten the image at each time step into a 1D vector to be processed by their fully connected layers, thereby discarding spatial information critical for accurate reconstruction of the output. 

To address this, ConvLSTM layers have been introduced \cite{shi2015convolutional}, which replace fully connected operations with convolutional ones, enabling the network to preserve spatial structure while modeling temporal dynamics. While promising, prior work employing ConvLSTM layers in the context of biomedical image analysis typically does so within relatively shallow CNN backbones \cite{almiahi2024novel}, which may limit the quality and robustness of the extracted features, particularly in heterogeneous or noisy imaging environments.

In this work, we propose ReCoGNet, a novel hybrid segmentation model that leverages both deep pretrained spatial feature extraction and sequential memory modeling. Specifically, we treat the MRI volume as a temporally ordered sequence, where each slice corresponds to a time step. A deep pretrained DeepLabV3 backbone \cite{chen2017rethinking} is used to extract rich, multiscale semantic features from each slice, and these features are then passed to a recurrent head composed of ConvLSTM layers. The recurrent head integrates contextual cues from preceding slices to inform segmentation predictions for the current slice, enabling smoother, more anatomically consistent delineations across the volume. The use of a pretrained backbone not only accelerates convergence but also enhances robustness, as it provides high-quality representations that generalize well across patients and imaging conditions.

We evaluate our proposed method on the PROMISE12 dataset \cite{litjens2014evaluation}, a widely used benchmark for prostate MRI segmentation. To emulate real-world clinical constraints, we conduct experiments under limited supervision using only 26 annotated volumes for training. In addition to standard test-time evaluations on clean data, we assess model robustness under simulated image degradation conditions—specifically, test volumes with reduced contrast in the second half of each MRI sequence. We compare our model against several state-of-the-art 2D and 3D segmentation baselines.

Our results demonstrate that the proposed model achieves superior segmentation performance across a range of metrics, including precision, recall, Intersection over Union (IoU), and Dice Similarity Coefficient (DSC), on both clean and degraded test sets. These findings highlight the advantages of combining deep spatial feature extraction with recurrent temporal modeling, and suggest that our approach is particularly well-suited for clinical deployment in settings with limited labeled data and variable imaging quality.

The rest of the paper is structured as follows: Section II presents the related work section. The methodology is described next, followed by a description of the experiments and the evaluation setup. The results are discussed next, followed by the conclusion.

\section{Related Work}
CNNs have been the dominant approach for biomedical image segmentation applications for the last several years. Notably, the U-Net architecture \cite{ronneberger2015u} has become a cornerstone architecture due to its symmetric encoder-decoder structure with skip connections, enabling effective feature propagation and precise segmentation outcomes. Over the years, several variants of UNet have been proposed such as the Attention UNet \cite{oktay2018attention} and the UNet++ \cite{zhou2018unet++}. 

The Attention UNet adds attention gates (AGs) to the standard U-Net. These gates are applied to the skip connections to suppress irrelevant features and emphasize informative regions in the encoder feature maps before concatenation with the decoder features. The UNet++ on the other hand aims to reduce the semantic gap between encoder and decoder feature maps through intermediate convolutional blocks, thereby improving segmentation performance.

While effective, these 2D CNNs operate on individual slices independently and do not capture contextual relationships between adjacent slices in 3D MRI data. To better utilize volumetric context, 3D CNNs such as VNet \cite{milletari2016v} have been applied to prostate MRI segmentation. These models process the entire MRI volume at once, thereby capturing spatial dependencies across slices. However, 3D CNNs are significantly more computationally demanding and prone to overfitting. Additionally, memory constraints often limit input resolution or batch size, which may degrade performance in fine-grained segmentation tasks. 

To preserve spatial structure while incorporating sequential memory, Convolutional LSTM (ConvLSTM) layers were introduced by Shi et al. \cite{shi2015convolutional}. ConvLSTMs replace fully connected operations with convolutional ones, enabling the network to retain spatial correlations over time. However, many existing implementations for biomedical imaging applications embed ConvLSTM layers within relatively shallow convolutional architectures, potentially limiting the quality of feature representations.

Recent studies have shown that using deep pretrained backbones, such as ResNet or DeepLabV3 \cite{chen2017rethinking}, can significantly improve segmentation performance and generalization in low-data regimes. These models, pretrained on large natural image datasets (e.g., ImageNet or COCO), capture rich semantic features and can be fine-tuned on target medical datasets with limited supervision. However, most prostate MRI segmentation studies employing pretrained networks focus solely on spatial modeling and do not leverage temporal dependencies across slices. 

To the best of our knowledge, our work is among the first to leverage the rich feature extraction capabilities of deep pretrained convolutional backbones along with the sequential modeling capablities of the ConvLSTM layers.

\section{Methodology}

In this section, we discuss the architecture of the deep network and provide a detailed explanation of the input data and how the model is trained.

\subsection{Pretrained DeepLabV3 Backbone}
The DeepLabV3 \cite{chen2017rethinking} architecture is a widely adopted deep learning model for semantic segmentation, designed to capture multi-scale contextual information while maintaining spatial accuracy in feature representations. DeepLabV3 employs an Atrous Spatial Pyramid Pooling (ASPP) module to effectively extract features at multiple receptive field scales, addressing the challenge of segmenting objects of varying sizes within an image. Unlike conventional convolutional neural networks (CNNs) that rely on downsampling operations such as max pooling and strided convolutions—potentially leading to loss of fine spatial details—DeepLabV3 incorporates atrous (dilated) convolutions, which enable feature extraction at larger receptive fields without increasing computational cost or reducing resolution.

At the core of DeepLabV3 is the ASPP module, which consists of parallel convolutional layers with different atrous rates. This structure allows the network to capture both local and global context by processing the input feature maps at multiple scales. The ASPP module is complemented by batch normalization layers and a global average pooling branch, further enhancing the model's ability to recognize structures at different levels of granularity. The final segmentation output is generated by fusing the multi-scale features extracted by the ASPP module and passing them through a lightweight decoder, which refines spatial details for improved segmentation accuracy.

\subsection{Recurrent Convolutional Head}
The ConvLSTM block, as defined in the implementation, consists of a convolutional recurrent unit that processes a sequence of MRI slices while maintaining hidden and cell states in a structured 2D format. Traditional recurrent neural networks (RNNs) and their variants, such as LSTMs and GRUs, are well-suited for modeling temporal dependencies in sequential data but operate on unstructured feature vectors, making them suboptimal for spatially correlated imaging data. In contrast, ConvLSTMs extend the memory-based processing of LSTMs to structured 2D feature maps by replacing fully connected transformations with convolutional operations, thereby preserving spatial information across MRI slices. At each time step, the input slice is concatenated with the previous hidden state and passed through a series of convolutional layers to compute the gating mechanisms—input gate, forget gate, output gate, and cell gate. These gates regulate information flow, allowing the model to retain relevant spatial-temporal features and discard redundant information. By maintaining structured feature maps in memory, the ConvLSTM enhances the network's ability to capture inter-slice dependencies, improving segmentation accuracy in volumetric medical imaging. For an input MRI sequence \( \mathbf{X} = \{X_1, X_2, ..., X_S\} \) consisting of \( S \) slices, where each slice \( X_t \in \mathbb{R}^{C \times H \times W} \) represents a 2D feature map with \( C \) channels, height \( H \), and width \( W \), the ConvLSTM cell maintains hidden states \( \mathbf{H}_t \) and cell states \( \mathbf{C}_t \) at each time step \( t \). The update equations for ConvLSTM at each time step are defined as follows:
\begin{align}
\mathbf{i}_t &= \sigma(\mathbf{W}_{xi} * \mathbf{X}_t + \mathbf{W}_{hi} * \mathbf{H}_{t-1} + \mathbf{b}_i) \\
\mathbf{f}_t &= \sigma(\mathbf{W}_{xf} * \mathbf{X}_t + \mathbf{W}_{hf} * \mathbf{H}_{t-1} + \mathbf{b}_f) \\
\mathbf{g}_t &= \tanh(\mathbf{W}_{xg} * \mathbf{X}_t + \mathbf{W}_{hg} * \mathbf{H}_{t-1} + \mathbf{b}_g) \\
\mathbf{o}_t &= \sigma(\mathbf{W}_{xo} * \mathbf{X}_t + \mathbf{W}_{ho} * \mathbf{H}_{t-1} + \mathbf{b}_o) \\
\mathbf{C}_t &= \mathbf{f}_t \odot \mathbf{C}_{t-1} + \mathbf{i}_t \odot \mathbf{g}_t \\
\mathbf{H}_t &= \mathbf{o}_t \odot \tanh(\mathbf{C}_t)
\end{align}
where \( \mathbf{i}_t, \mathbf{f}_t, \mathbf{o}_t \in \mathbb{R}^{C \times H \times W} \) are the input, forget, and output gates, respectively, \( \mathbf{g}_t \in \mathbb{R}^{C \times H \times W} \) represents the candidate memory cell update, \( \mathbf{C}_t \) is the cell state at time \( t \), which stores long-term dependencies, \( \mathbf{H}_t \) is the hidden state at time \( t \), which is passed to the next time step and used in the segmentation output, \( \sigma(\cdot) \) denotes the sigmoid activation function, \( \tanh(\cdot) \) is the hyperbolic tangent function, \( * \) represents a convolution operation, and \( \odot \) denotes element-wise multiplication.
The overall architecture follows an encoder-decoder design, where the encoder extracts hierarchical feature representations using alternating convolutional and ConvLSTM layers. The encoder begins with standard convolutional layers to learn spatial feature embeddings, followed by ConvLSTM layers that model slice-to-slice relationships. This hybrid approach allows the network to leverage both spatial and temporal information for improved segmentation. The decoder mirrors the encoder structure, employing deconvolutional layers for upsampling and additional ConvLSTM layers to refine feature maps before producing the final segmentation mask.

\subsection{Loss Function}
For an input MRI sequence \( \mathbf{X} = \{X_1, X_2, ..., X_S\} \) consisting of \( S \) slices, the pretrained DeepLabV3 backbone extracts per-slice feature embeddings. Let the function \(f_{DL}(X_t;\theta_{DL})\) denote the feature extraction operation. This is then represented as 
\begin{equation}
    F_{t} = f_{DL}(X_t;\theta_{DL}), \quad t=1,...S
    \label{eq:feat_extraction}
\end{equation}
where \( X_t \in \mathbb{R}^{C_{\text{in}} \times H \times W} \) is the input MRI slice at time \( t \), \( C_{\text{in}} \) is the number of input channels, and \( H, W \) are the height and width of the image, respectively. \(\theta_{\text{DL}}\) denotes the trainable parameters of the DeepLabV3 backbone, and \( F_t \in \mathbb{R}^{C_{\text{feat}} \times H \times W} \) is the extracted feature representation, where \( C_{\text{feat}} \) is the number of output feature channels. Applying this to the entire image sequence generates the feature embedding sequence \( \mathbf{F}=\{F_1, F_2, ..., F_S\}\). The Recurrent Convolutional head, represented by \(f_{RC}(.)\) thereafter ingests this sequence of feature activations to generate the raw logit maps for the prostate for each slice in the sequence. This is represented as 
\begin{equation}
    H_{t} = f_{RC}(X_t,X_{t-1},...X_{1};\theta_{RC}), \quad t=1,...S
    \label{eq:convlstm}
\end{equation}
where \( H_t \in \mathbb{R}^{1 \times H \times W} \) is the prostate logit map at time slice \( t \), and \( H, W \) are the height and width of the image, respectively. \(\theta_{\text{RC}}\) denotes the trainable parameters of the recurrent convolutional head.

A weighted cross entropy loss is used to measure the discrepancy between the predicted segmentation outputs and the ground-truth annotations \( \mathbf{Y} = \{Y_1, Y_2, ..., Y_S\} \), where \(Y_t\in \{0,1\}^{H\times W}\). The total loss is computed as 

\begin{equation}
    \text{loss} = \sum_{i=1}^{N} \text{BCE}(\mathbf{H_{i}}, \mathbf{Y_{i}}),
    \label{eq:loss}
\end{equation}

where \(\mathbf{H_i}\) and \(\mathbf{Y_i}\) denote the predicted and ground-truth segmentations for example $i$, BCE refers to the application of the binary cross-entropy loss, and $N$ refers to the total number of training examples in the dataset.

\section{Experiments}

\subsection{Dataset Details}
The PROMISE12 \cite{litjens2014evaluation} dataset was made available for the MICCAI 2012 prostate segmentation challenge. Transversal T2-weighted Magnetic Resonance (MR) images of 50 patients with various diseases were acquired at different locations with several MRI vendors and scanning protocols. These cases include a transverse T2-weighted MR image of the prostate. The training set comprises a representative collection of MR images typically acquired in clinical settings. The data is sourced from multiple centers and vendors, featuring various acquisition protocols (e.g., differences in slice thickness and use of endorectal coil). The set is chosen to ensure a diversity of prostate sizes and appearances. Each case in the training set includes a corresponding reference segmentation. We randomly selected 26 patients to be used for training various network architectures. The other 24 are used for testing the generalization performance of the trained models. 

\subsection{Benchmark Models}
To compare the performance of our model to existing literature, we select the following approaches: UNet, Attention UNet, UNet++, VNet, and DeepLabV3. The first three models are selected for their simplicity, small size, and reputation as high performing 2D convolutional segmentation models. The VNet model is selected to represent 3D convolutional segmentation models. Finally, the DeepLabV3 model is chosen for its sophisticated, multiscale feature extraction capabilities using its deep, pretrained backbone comprising several residual blocks. Since the proposed network architecture uses DeepLabV3 as its backbone, the latter would serve as a good baseline to properly account for the effect of the recurrent convolutional head.   

\subsection{Evaluation Setup}
We conduct our evaluation of the proposed model and the baseline approaches under two configurations of the test set. In the first configuration, we use the original, clean images from the patient MRI data. Under the second evaluation setting, the MRI data from each patient is corrupted so that the first half of a given MRI sequence appears noise-free but the second half has its contrast reduced by 20\%. This serves two purposes: first, it measures the robustness of the various approaches to noisy MRI images; secondly, it evaluates the recurrent convolutional head in the proposed network architecture in terms of how well it can predict on a suddenly degraded image slice by relying on the memory its past inputs. It is worth mentioning that the same set of trained models is used to perform inference in both configurations. In other words, the models are trained on clean image data and later evaluated on both clean and corrupted MRI sequences. For training details, please refer to the Appendix. 

\section{Results and Discussion}
Table \ref{tab:prostate_seg_results} shows the segmentation results obtained through various methods on both the clean and corrupted versions of the test set in terms of the precision, recall, F1 score, and dice similarity coefficient, respectively. The results have been averaged over all examples in the test set. 

For the clean test set, the proposed method outperforms other baseline methods used in the study, although the difference is relatively minor compared to the DeepLabV3. This is because both methods leverage the deep pretrained backbone of the same underlying architecture. Under clean imaging conditions, DeepLabV3 is able to localize the prostate well even in the absence of the recurrent convolutional head. 

On the corrupted test set, all baseline methods suffer significant performance degradation, as evidenced by the various evaluation metrics. However, the proposed method demonstrates a high degree of robustness, maintaining significantly higher levels of DSC and F1 scores compared to second best-performing method: DeepLabV3. This observation can largely be attributed to the recurrent convolutional head in the proposed network architecture that is able to use knowledge from seeing past inputs to make more informed predictions for future slices in a given MRI sequence. 

Also note that the VNet model largely fails to segment the prostate owing to the limited number of labeled examples available for training the 3D convolutional kernels present in the architecture. It is also worth noting that the UNet, Attention UNet, and UNet++ methods do not perform well either despite all comprising only 2D convolutional kernels, underscoring the importance of having deep pretrained backbone networks for robust and generalizable feature extraction from raw images.

\begin{figure*}
    \centering
    \includegraphics[width=\linewidth]{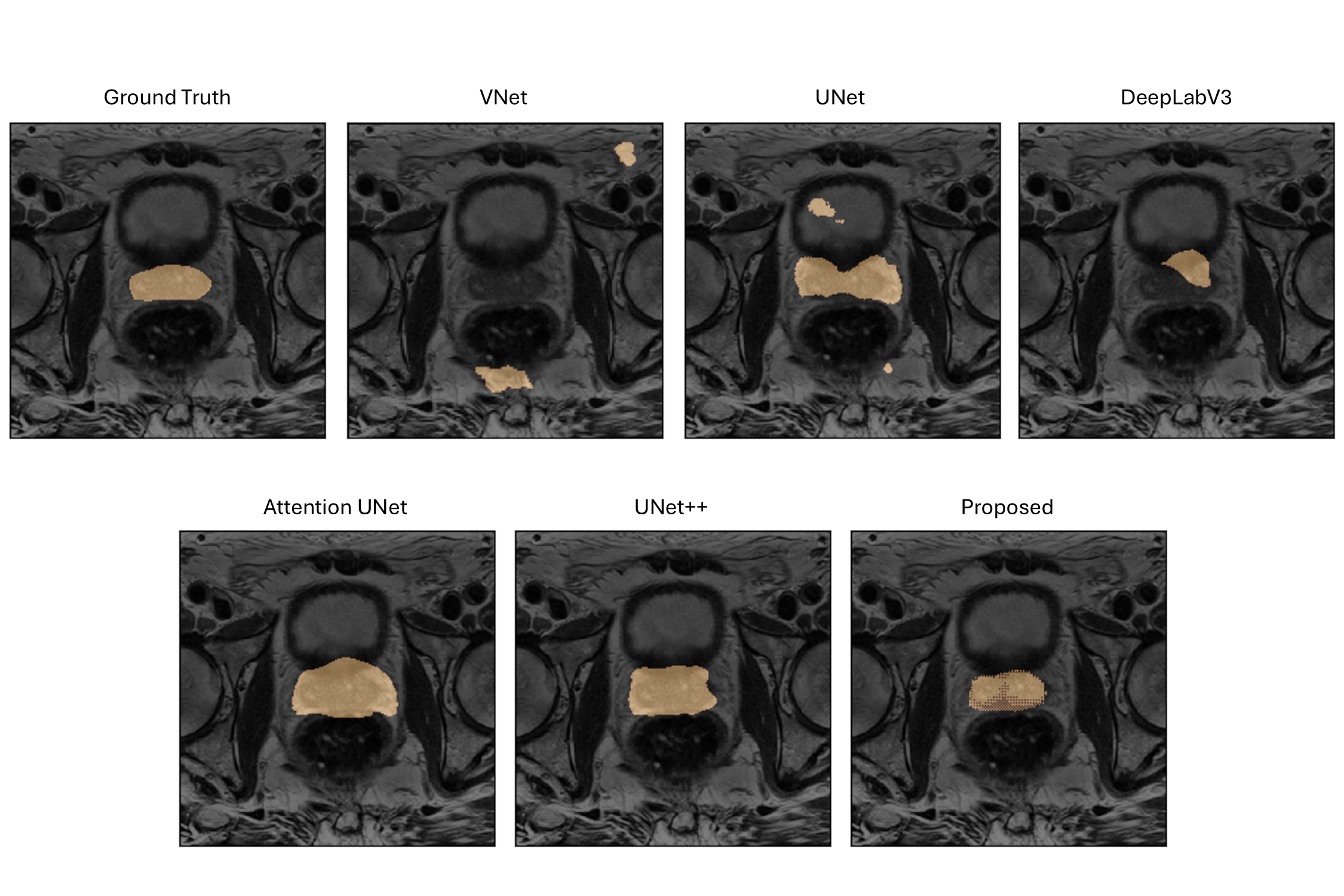}
    \caption{Comparison of segmentation results across various methods compared to the ground-truth (top left). The proposed method (bottom right) can be observed to capture the prostate gland most faithfully compared to benchmark methods.}
    \label{fig:visual}
\end{figure*}

Fig. \ref{fig:visual} shows a qualitative comparison of the segmentation outputs on an MRI image slice taken from a randomly selected test patient along with the reference annotations. It can be observed that the proposed method produces the most visually similar output compared to the ground-truth segmentation. Other methods exhibit considerable deviation from the ground-truth as observed in the shape, position, and size characteristics of their segmentation outputs. 

\begin{figure*}
    \centering
    \includegraphics[width=\linewidth]{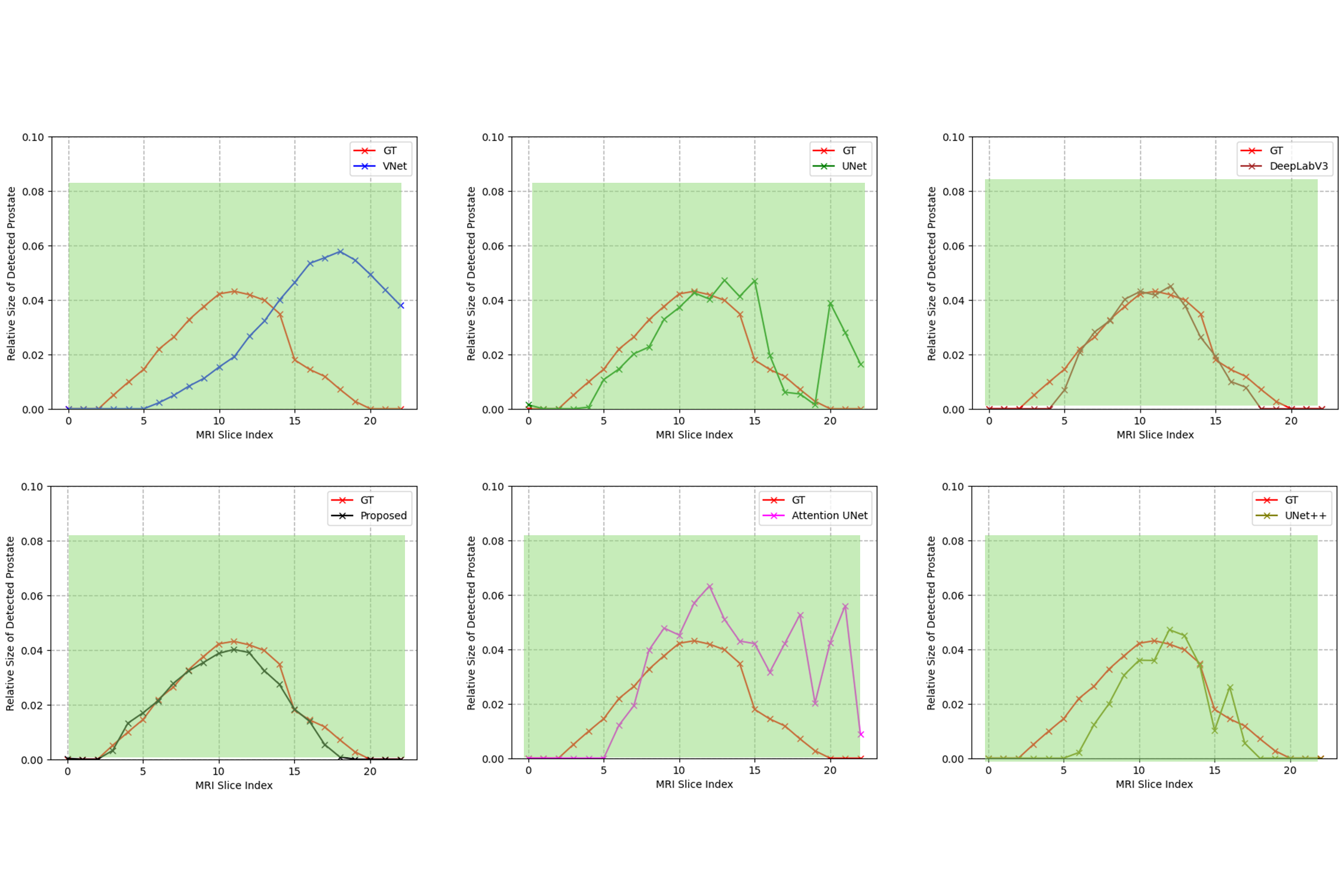}
    \caption{Comparison of relative prostate sizes (normalized by the image dimensions) across segmentation methods for each MRI slice from a single patient under noise-free imaging conditions (depicted by the green color in the background). Ground-truth measurements are shown alongside predictions from the proposed model and baseline methods from the literature. The predictions of the proposed model (first plot in the second row) can be observed to be best aligned to the reference annotations.}
    \label{fig:patient_4_profiles}
\end{figure*}

We further investigated how well each method can localize the prostate gland on a complete MRI sequence compared to the reference segmentation. Towards this end, a patient in the test set was randomly selected and used to produce segmentation outputs from all methods used in the study. Thereafter, we plotted the relative size of the detected prostate for each slice in the sequence, along with that of the ground-truth annotation. These plots can be seen in Fig. \ref{fig:patient_4_profiles}. The proposed method is seen to mimic the ground-truth prostate size profile most faithfully compared to other baseline approaches. 

\begin{figure*}
    \centering
    \includegraphics[width=\linewidth]{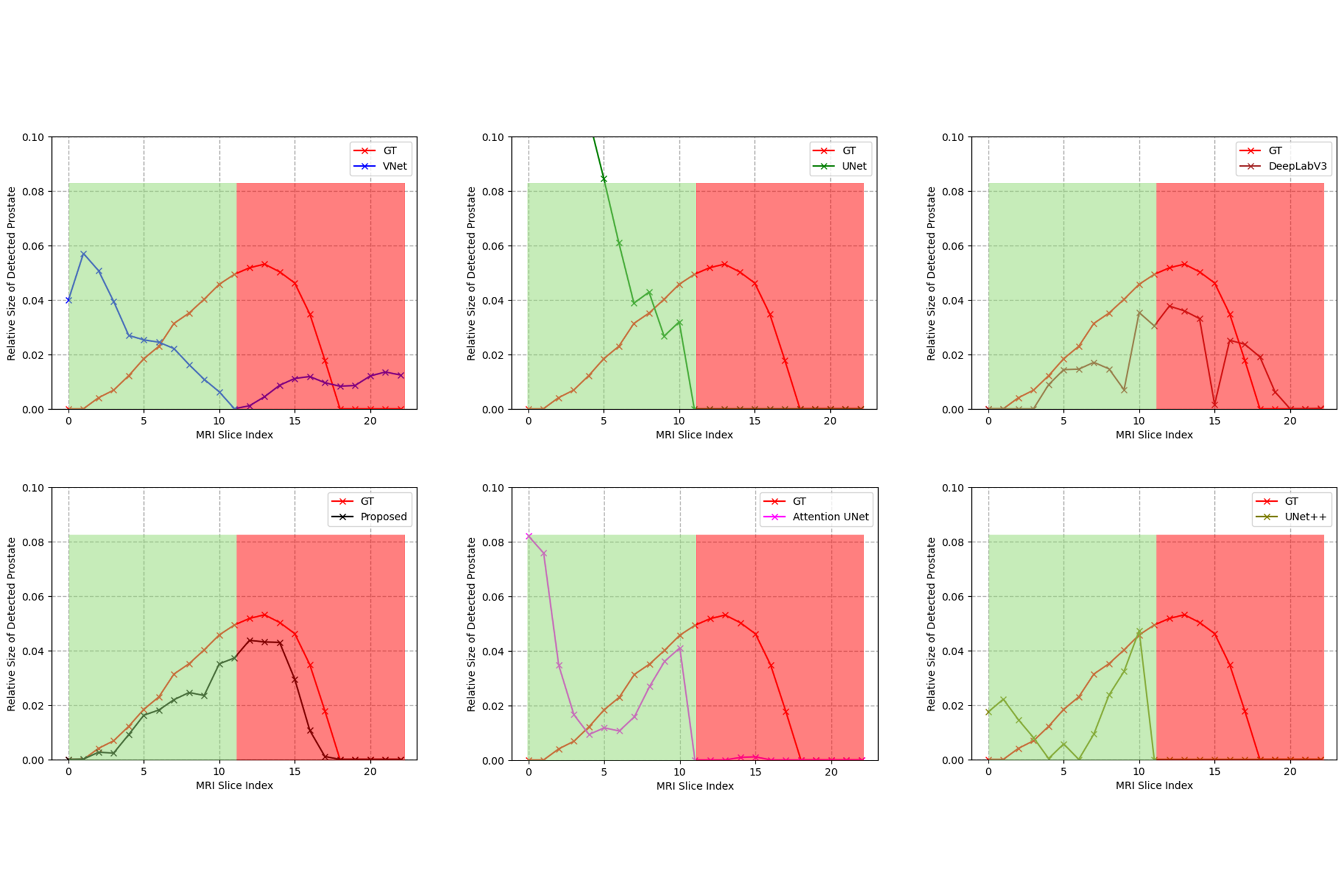}
    \caption{Comparison of relative prostate sizes (normalized by the image dimensions) across segmentation methods for each MRI slice from a single patient under noisy imaging conditions. The first half of the sequence (denoted by green) consists of the original noise free images while the second half simulates low contrast imaging into the image slices (denoted by red). Ground-truth measurements are shown alongside predictions from the proposed model and baseline methods from the literature. The proposed model's predictions (first plot on the second row) can be observed to be best aligned to the reference annotations.}
    \label{fig:patient_2_profiles}
\end{figure*}

We reproduced the same experiment for a different random test patient, this time with corrupted imaging conditions. This is shown in Fig. \ref{fig:patient_2_profiles}. Where other methods can be observed to demonstrate significant performance deterioration, the proposed model can still capture the underlying prostate profile to a high degree of accuracy.

% \begin{table}[h]
%     \centering
%     \renewcommand{\arraystretch}{1.3} % Increase row height
%     \begin{tabular}{lcccc}
%         \toprule
%         \textbf{Architecture} & \textbf{Precision} & \textbf{Recall} & \textbf{F1-score} & \textbf{DSC} \\
%         \midrule
%         \multicolumn{5}{c}{\textbf{Baseline Architectures}} \\
%         UNet & 0.39 & 0.76 & 0.51 & 0.49 \\
%         UNet++ & 0.61 & 0.82 & 0.70 & 0.66\\
%         DeepLabV3 & 0.80 & 0.78 & 0.79 & 0.78 \\
%         Attention UNet & 0.39 & 0.84 & 0.53 & 0.52 \\
%         UNet3D & 0.33 & 0.46 & 0.38 & 0.37 \\
%         \midrule
%         \multicolumn{5}{c}{\textbf{Baseline + Proposed Head}} \\
%         UNet + Proposed Head & 0.39 & 0.73 & 0.51 & 0.49 \\
%         UNet++ + Proposed Head & 0.47 & 0.90 & 0.62 & 0.60 \\
%         DeepLabV3 + Proposed Head & 0.84 & 0.81 & 0.83 & 0.82 \\
%         Attention UNet + Proposed Head & 0.42 & 0.82 & 0.55 & 0.54 \\
%         \bottomrule
%     \end{tabular}
%     \vspace{2mm}
%     \caption{Segmentation performance of different architectures on the benchmark dataset.}
%     \label{tab:prostate_seg_results}
% \end{table}

\begin{figure*}
     \centering
    \includegraphics[width=\linewidth]{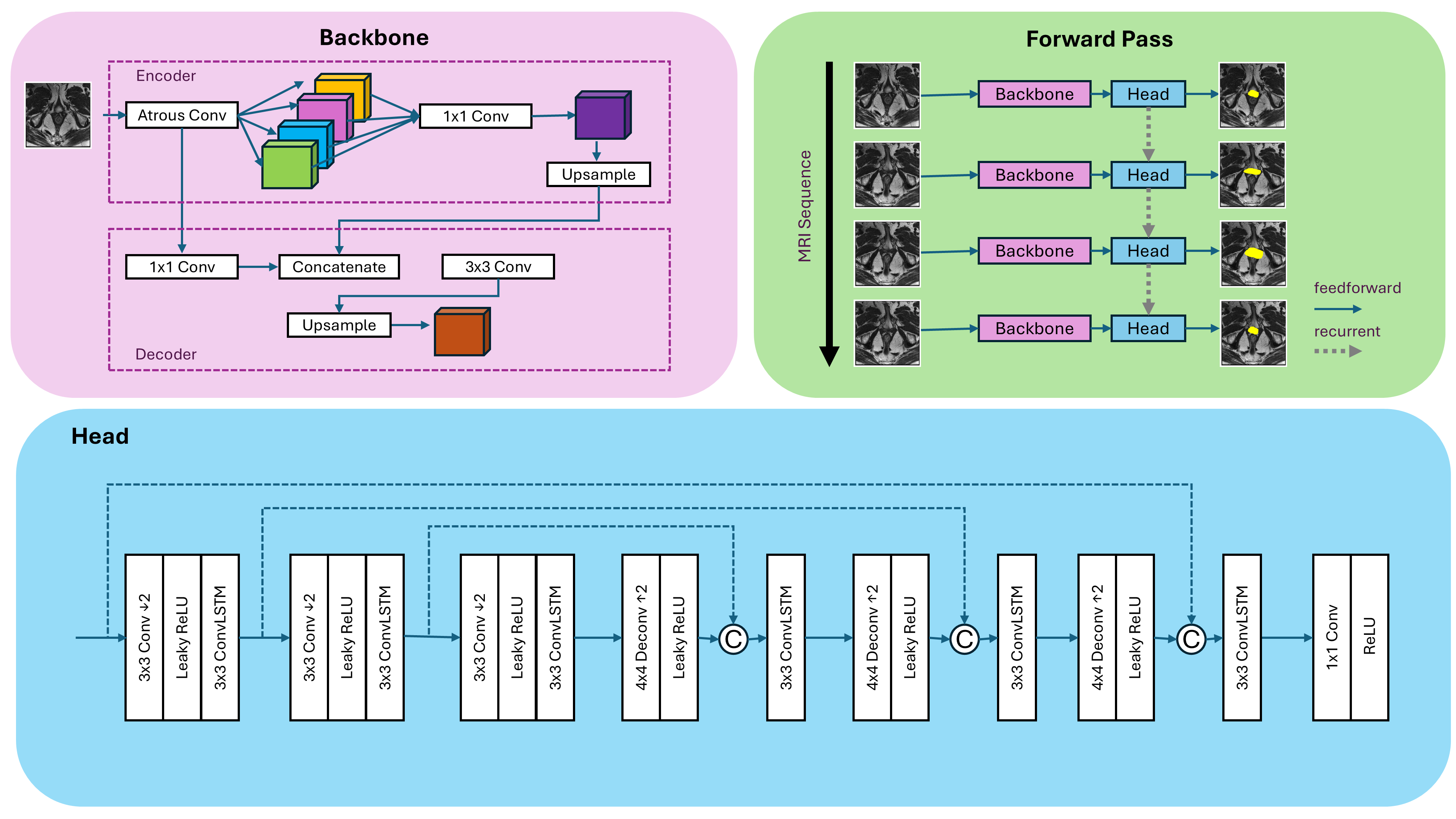}
    \caption{Figure shows the inner workings of the proposed model architecture. The backbone consists of DeepLabV3 extracting and aggregating multiscale features from raw MRI slices to produce high level feature maps. For each MRI slice, the backbone's output is further processed by the head to produce the final segmentation map. The head retains memory of past inputs in the activations of its recurrent convolutional layers and uses them in conjunction with feature maps from the current slice to generate its segmentation output. The head consists of regular convolutional layers and recurrent convolutional layers organized in an encoder-decoder configuration. Skip connections from the encoder send its information to the decoder to mitigate the loss of feature resolution in the encoder.}
    \label{fig:architecture}
\end{figure*}

\begin{table*}[h]
    \centering
    \renewcommand{\arraystretch}{1.3} % Increase row height
    \begin{tabular}{lcccc|cccc}
        \toprule
        \textbf{Architecture} & \textbf{Precision} & \textbf{Recall} & \textbf{F1-score} & \textbf{DSC} & \textbf{Precision} & \textbf{Recall} & \textbf{F1-score} & \textbf{DSC} \\
        \midrule
        & \multicolumn{4}{c}{\textbf{Clean Test Set}} & \multicolumn{4}{c}{\textbf{Corrupted Test Set}}\\
        UNet \cite{ronneberger2015u} & 0.39 & 0.76 & 0.51 & 0.49 & 0.04 & 0.12 & 0.06 & 0.06 \\
        UNet++ \cite{zhou2018unet++} & 0.61 & 0.82 & 0.70 & 0.66 & 0.54 & 0.56 & 0.55 & 0.52\\
        Attention UNet \cite{oktay2018attention}& 0.39 & 0.84 & 0.53 & 0.52 & 0.30 & 0.52 & 0.38 & 0.36\\
        VNet \cite{milletari2016v} & 0.33 & 0.46 & 0.38 & 0.37 & 0.24 & 0.28 & 0.26 & 0.24 \\
        DeepLabV3 \cite{chen2017rethinking} & 0.80 & 0.78 & 0.79 & 0.78 & 0.84 & 0.63 & 0.72 & 0.70\\
        \textbf{Proposed} & \textbf{0.84} & 0.81 & \textbf{0.83} & \textbf{0.82} & \textbf{0.84} & \textbf{0.75} & \textbf{0.79} & \textbf{0.78}\\
        \bottomrule
    \end{tabular}
    \vspace{2mm}
    \caption{Segmentation performance of different architectures on the benchmark dataset.}
    \label{tab:prostate_seg_results}
\end{table*}

\section{Conclusion}
We presented an approach for automatic prostate gland segmentation in T2-weighted MRI sequences leveraging a deep, pretrained segmentation backbone attached to a head comprising recurrent convolutional layers. Our experiments demonstrate that the deep convolutional backbone from the popular DeepLabV3 segmentation model produces robust, generalizable features that are further refined by the recurrent convolutional head to model memory of past input slices in MRI sequences. We benchmarked our approach against a variety of baseline methods in the literature, including the traditional UNet and its various 2D/3D variants. Our proposed method is seen to significantly outperform baseline approaches in limited label settings. Further, our method exhibits considerable resilience to sudden degradations in imaging conditions such as changes in contrast compared to other approaches. Given its robustness to noisy inputs and a high level of generalization in limited label settings, our method shows significant potential for adoption in clinical workflows involving MRI-based prostate localization.

\ifCLASSOPTIONcaptionsoff
  \newpage
\fi



\bibliographystyle{IEEEtran}
\bibliography{bibtex/bib/refs}

\appendix
\subsection*{Additional Training Details}
\label{app:training}
All models are trained in PyTorch for 100 epochs with a fixed learning rate of 0.001 using the ADAM optimizer. The MRI sequences in the PROMISE12 dataset exhibit varying sizes. During training and inference, all images are resampled to 224 $\times$ 224 size. Image amplitudes are scaled to lie between 0 and 1. To improve image contrast, histogram equalization is carried out over all slices in the MRI data. 

To keep training costs tractable, we randomly select a subsequence consisting of 10 consecutive slices from the full MRI sequence data of any given patient. For DeepLabV3, UNet, Attention UNet, and UNet++, these slices are ordered in the batch dimension. For the proposed model, these image slices are processed image-wise by the backbone. Thereafter, the output of the backbone is rearranged so that the respective feature maps for each image slice form a sequence to be input to the recurrent convolutional head. Since inference requires less GPU VRAM, each MRI sequence is processed as a whole (versus subsamples of size 10 as in training). For VNet, each subsequence is padded to a standard size of 96 in the sequence length dimension to allow processing by the full depth of the model. However, the loss is only computed for the first 10 slices to prevent padded slices from contributing to the backpropagated gradients.   Both training and inference is carried out on a single NVIDIA T4 GPU featuring 16GB of GPU VRAM.


% \end{IEEEbiography}


% % insert where needed to balance the two columns on the last page with
% %\newpage



%\vfill


\end{document}